\documentclass[12pt]{article}
\usepackage{graphicx}
\def\@oddhead{}\def\@evenhead{}
\def\@oddfoot{}
\def\@evenfoot{\@oddfoot}
\begin{document}
\newcommand{\ra}{\rightarrow}
\begin{flushright}
{\bf ATLAS Internal Note \\
PHYS-No-30\\
7 September 1993}
\end{flushright}
\vskip 0.7cm
\begin{center}
{\large\bf {On a possibility to extract a signal \\ 
from heavy gluino cascade decays via isolated muon detection \\
at the LHC energies}}\\[7mm]
{\normalsize\it {V.\,I. Klyukhin, A.\,A. Neushkin, A.\,A. Sokolov}}\\[3mm]
{\normalsize {SINP, Lomonosov Moscow State University, Moscow, Russia \\
E-mail: Vyacheslav.Klyukhin@cern.ch \\
Institute for High Energy Physics, Protvino, Moscow reg., Russia \\
E-mail: sokolov\_a@ihep.ru}}\\[4mm]
\end{center}
\vspace{0.7cm}
\begin{center}
\begin{minipage}{14cm}
\small

\begin{center}
{\bf Abstract}
\end{center}

800~GeV gluino pair production is simulated using ISASUSY~1.1 MC-program. 
The cascade decay of the  gluino into two lightest  neutralinos, 
$\tilde{g} \ra \tilde{\chi}_2^0 X$, $\tilde{\chi}_2^0 \ra \mu^+ \mu^-
\tilde{\chi}_1^0$, is studied for 
a convenient set of SUSY parameters: $\mu = -200$~GeV, $\tan{\beta} = 2$, 
average squark masses of 1600~GeV, charged Higgs mass of 500~GeV, and
top quark mass of 140~GeV. With such parameters the mass difference between
$\tilde{\chi}^0_2$ and $\tilde{\chi}_1^0$ is less than the mass of $Z$. As a
background, $t \bar t$, $Z + jets$, $WW$, $WZ$ and $ZZ$ events were generated
using ISAJET~6.50 MC-program. The granularities and energy resolutions of the
calorimeters in the pseudorapidity range of $|\eta | < 5$ are taken into 
account. If missing transverse momentum is greater than 200~GeV, the number of
jets in the calorimeters is 4 or more, the invariant
mass of two isolated muons is less than 80~GeV, the background
in the muon transverse momentum distribution is negligible.
\end{minipage}
\end{center}
\vspace{0.7cm}
\section*{\large\bf Introduction}

The purpose of this paper is to investigate whether it is possible to extract 
a signal from the heavy gluino decay via the cascade chain 

\begin{equation}
\tilde{g} \ra \tilde{\chi}_2^0 q\bar q, \quad \tilde{g} \ra \tilde{\chi}_2^0 g,
\end{equation}
\begin{equation}
\tilde{\chi}_2^0 \ra Z^* \tilde{\chi}_1^0 \ra l^+ l^- \tilde{\chi}_1^0
\quad (l=e,\mu )
\end{equation}

\noindent for the standard SUSY parameters used for the gluino production
studies in the ATLAS Letter of Intent~\cite{ATLAS}. An interest in these 
cascade decays is caused by the previous investigations~\cite{Sokolov} in which
a possibility to estimate the mass of $\tilde{\chi}_2^0$ by measuring the
transverse momentum distribution of the muons coming from the decay (2) was
demonstrated at the UNK energies. 
Using MC-program ISAJET~\cite{ISAJET} for their simulations, the
authors of~\cite{Sokolov} showed that the position of the maximum in the muon
$p_T$-distribution reflects the relationship between the masses of 
$\tilde{\chi}_2^0$ and $\tilde{\chi}_1^0$ as follow: 

\begin{equation}
p_{T_{\mu}}^{max} = \frac{m_{\tilde{\chi}_2^0}^2 - 
m_{\tilde{\chi}_1^0}^2}{4m_{\tilde{\chi}_2^0}}
\end{equation}

In the present paper we repeate those studies at the LHC energy using the
new MC-program ISASUSY 1.1~\cite{ISASUSY} to generate the signal from gluino
pair production, and ISAJET 6.50~\cite{ISAJET} to simulate all backgrounds
to this signal. We pay special attention to such parameters of
the central and forward calorimeters as granularity and energy resolution.

ISASUSY 1.1 was modified to generate not only the signal from the decays of
$\tilde{g} \ra X$, but also from the forced decays (1)-(2) directly. As
backgrounds to these processes we consider the production of $t\bar t$, 
$Z + jets$ and $WW$, $WZ$, $ZZ$ continuum at high transverse momenta with
subsequent decays into leptons (i.e. $t \ra Wb$, $W \ra l\nu$, $b \ra X$, 
$Z \ra ll$, where $l = \mu , \tau$ and $\tau \ra \mu \nu \nu$).

\section*{\large\bf The choice of SUSY parameters}

The present status of the minimal SUSY extension of the Standard Model 
(MSSM)~\cite{MSSM} gives us a set of model parameters as follow:

\begin{itemize}
\item the gluino mass, $m_{\tilde{g}}$, and an average squark mass, 
$m_{\tilde{q}}$;

\item the Higgs-higgsino mass term, $\mu$;

\item the ratio of the vacuum expectation values of two Higgs doublets required
by the MSSM, $tan{\beta}$;

\item the mass of the charged Higgs particle, $m_{H^+}$.
\end{itemize}

In this paper we use the following set of SUSY parameters at the c.m. energy
of 16~TeV:
$m_{\tilde{g}} = 800$~GeV, $m_{\tilde{q}} = 2 m_{\tilde{g}}$, $\mu = -200$~GeV, 
$tan{\beta} = 2$, and $m_{H^+} = 500$~GeV. In addition to this set of
parameters we input the value of the top quark mass, the last unknown parameter 
of the Standard Model, $m_t = 140$~GeV. For those parameters the calculations of
the sparticle masses and branching ratios performed in the framework of ISASUSY
1.1 show that the mass difference of two lightest neutralinos,
$m_{\tilde{\chi}_2^0} -  m_{\tilde{\chi}_1^0}$,
is less than the mass of the real $Z$. Moreover, this difference remains less 
than the real $Z$ mass in the wide range of $m_{\tilde{g}}$ and $\mu$, as shown
in Figs.~1-9. This leads to a natural wish to try to extract a signal from the
cascade decays (1)-(2) of heavy gluinos at the chosen
set of SUSY parameters. 

As Figs.~10-15 display, the ISASUSY 1.1 gives branching ratios for
the decays (1) and (2) quite acceptable to detect
the leptons coming from the decays (2). Further we will consider the decay mode
of $\tilde{\chi}_2^0 \ra \mu^+ \mu^- \tilde{\chi}_1^0$ as the basic one, but all
our calculations are valid for both leptons mentioned above. 

\section*{\large\bf The calorimeter simulation}

We assume the R-parity conservation, and that $\tilde{\chi}_1^0$ is the 
lightest supersymmetric particle (LSP) which escapes the apparatus. This 
causes high missing transverse momentum in the events with heavy gluino 
pairs. It is of interest to study the influence of the calorimeter
performance to this kinematic variable. 

To simulate the calorimeter, particle energies were deposited in the grids with
different energy smearings and different granularities in ($\eta ,\phi$)-space,
depending on the $\eta$ coverage, where $\eta$ and $\phi$ denote the
pseudorapidity and azimuthal angle. In the central region, $|\eta |<3$, we
used a cell size of $\Delta \eta \times \Delta \phi = 0.1 \times 0.1$ for both
electromagnetic and hadronic central calorimeters. Their energy resolutions 
were taken as  $(\Delta E/E)_{em} = 10\%/\sqrt{E} \oplus 1\%$ and 
$(\Delta E/E)_{had} = 50\%/\sqrt{E} \oplus 3\%$, respectively. All neutrinos 
and LSP's, and also muons and electrons originated from W, Z and sparticles
(including intermediate $\tau$ decays) were not detected in the
central calorimeters. The transverse momenta of such electrons and
muons were then added to the transverse energies of the calorimeter cells,
$E_T$, to determine the missing transverse momentum from the calorimeter
measurements. The muon and electron momentum resolutions were not taken into
account.

In the backward-forward region, $3<|\eta |<5$, we used the granularity of
$0.2 \times 0.2$ and energy smearings of 
$(\Delta E/E)_{em} = 60\%/\sqrt{E} \oplus 1000\%/E \oplus 30\%$ and
$(\Delta E/E)_{had} = 150\%/\sqrt{E} \oplus 1000\%/E \oplus 7\%$. In this case
all neutrinos and LSP's, and also muons from W, Z and sparticle
decays were removed 
from the forward calorimeter, but not electrons, whose energies were smeared 
inside the cells in accordance with the chosen electromagnetic resolution. The 
transverse momenta of muons were not added to the transverse energies of 
calorimeter cells, taking into account the muon system coverage ($|\eta|<3$).

We specially used the worst case of hadronic and especially electromagnetic
resolutions to simulate the backward-forward region in order to look at the
influence on the missing transverse momentum resolution 
from these effects. Figs.~16-19
demonstrate the $p_T^{miss}$ for the signal and background
events with two opposite sign isolated muons for two different $\eta$
coverages (the criterion of the muon isolation is described below). Here and
below we will indicate the distributions in the Figures as follow:

\begin{itemize}
\item by {\em thin solid line} -- the events from $\tilde{g} \ra X$ decays;

\item by {\em thick solid line} -- the events from the forced (1)-(2) decays;

\item by {\em dashed line} -- the events from $t\bar t$ background;

\item by {\em dot-dashed line} -- the events from Drell-Yan $Z + jets$ 
production;

\item by {\em dotted line} --  the events from $WW$, $WZ$ and $ZZ$ continuum.
\end{itemize}

In Fig.~16 the value of missing transverse momentum is determined by sum of 
all stable charge particle and $\gamma$-ray transverse momenta over all 
$\eta$ range. In Fig.~17 the same is done for $|\eta |<5$. The $p_T^{miss}$
distribution in Fig. 18 is obtained from the sum of cell $E_T$ in the central
calorimeters only, and from the sum of the transverse momenta of charged leptons
not deposited in these calorimeters. Finally, Fig.~19 shows the $p_T^{miss}$
distribution obtained from the sum of cell $E_T$ in the range of $|\eta |<5$
and the sum of $p_T$ of charged leptons in the central region. 
One can see that Figs.~19 and 17 are very similar, so the
calorimeter performance does not influence the missing $p_T$
measurements much neither for the signal nor for the backgrounds except for $Z +
jets$ production.  

The second important signature of the signal events is a presence of several
quark jets with high transverse energy. A standard jet algorithm found 
the cell with the highest $E_T$ (larger than 3~GeV), used this
cell to define the axis of jet and collected all the energy in a cone with a
size of $\Delta R = \sqrt{(\Delta \eta )^2 + (\Delta \phi)^2} = 0.7$ to 
estimate the jet energy. Only cells with $E_T>1$~GeV were considered in this
procedure. If the transverse energy of the jet was lager then 100~GeV, the jet
was retained as a calorimeter jet. Fig.~20 shows the distribution of the number
of calorimeter jets with $E_T>100$~GeV in the signal and background events with
two opposite sign isolated muons.

\section*{\large\bf The cuts applied to the signal and background events}

Now we are in a position to describe the cuts applied to the generated 
events to reduce the backgrounds with respect to the signal. We generated two
kinds of the signal events: a sample of events with two gluinos decaying via 
all possible channels, and a sample of events in which one gluino decays via 
the cascade chain (1)-(2). The cross-sections and the numbers of such events 
corresponded to one LHC year are presented in the first two columns of Table~1.
The other columns of Table~1 relate to the generated background events caused 
by standard physics processes. To efficiently generate sufficient background
process statistics, we used the partons cuts, shown in brackets in Table~1.

\begin{center}
\small

{\bf Table~1} \\
\smallskip
Accumulated cut efficiencies for the signal and backgrounds \\
\smallskip
\begin{tabular}{|l|c|c|c|c|c|c|c|}    \hline

 & $\tilde{g} \ra X$ & $\tilde{g} \ra \tilde{\chi}_2^0 X$ &
   $t\bar t$ & $Z + jets$ & $WW$ & $WZ$ & $ZZ\dagger $ \\ \cline{4-8}
 &                   & $\ra \mu \mu \tilde{\chi}_1^0 X$ &
  \multicolumn{2}{c|}{($p_T^{top,Z} >200$~GeV)} &
  \multicolumn{3}{c|}{($p_T^{W,Z} >100$~GeV)} \\ \hline
 Production cross-section &   &     &        &        &      &     & \\
 ($pb$)         & 2.41 & 0.0121 &   302 &   76.9 & 5.93 & 2.51& 1.13 \\ \hline
 Events per $10^5~pb^{-1}$ &  &     &        &        &      &     & \\
 (with at least 2 muons &241000& 977 & 417330 & 214390 & 7347 & 866 & 1240 \\
 in backgrounds)$\ddagger $ & & &        &        &      &     & \\ \hline 
 Events passing      &        &     &        &        &      &     & \\
 succesive cuts:     &        &     &        &        &      &     & \\
 -- 2 isolated muons with& $\frac{2272}{0.943\%}$ & $\frac{333}{34.1\%}$ & 
 $\frac{150838}{36.1\%}$ & $\frac{158358}{73.9\%}$ & $\frac{4960}{67.5\%}$ &
 $\frac{195}{22.5\%}$ & $\frac{923}{74.5\%}$ \\
 ~~ $p_T > 7$~GeV    &        &     &        &        &      &     & \\
 -- $p_T^{miss} >200$~GeV & $\frac{1132}{0.47\%}$ & $\frac{166}{17.0\%}$ & 
 $\frac{3898}{0.934\%}$ & $\frac{859}{0.401\%}$ & $\frac{45}{0.612\%}$ &
 $\frac{13}{1.5\%}$ & $\frac{127}{10.2\%}$ \\
                     &        &     &        &        &      &     & \\
 -- $n_j > 3$        & $\frac{408}{0.169\%}$ & $\frac{79}{8.09\%}$ & 
 $\frac{34}{0.00815\%}$ & $\frac{9}{0.0042\%}$ & --- & --- & --- \\
                     &        &     &        &        &      &     & \\
 -- $m_{\mu \mu} <80$~GeV & $\frac{208}{0.0863\%}$ & $\frac{76}{7.78\%}$ & 
 $\frac{14}{0.00335\%}$ & $\frac{9}{0.0042\%}$ & --- & --- & --- \\
                     &        &     &        &        &      &     & \\ \hline
\end{tabular}

\smallskip
$\dagger $ Only ($Z \ra \mu \mu$)($Z \ra \nu \nu$) pair decays are forced for
this channel

$\ddagger $ An efficiency of $90\%$ is assumed for muon identification\\
\end{center}

\normalsize

One should note that the number of events in the second sample consists of
only 0.4\% of the full number of pair gluino events. Since we concentrate on
the decays (1)-(2), we expect a large background to these channels from other
gluino decays giving muons in the final state (e.g. the decays of 
$\tilde{g} \ra tX$ produce 
the muons with $p_T$-distribution close to that one from the decays (1)-(2)).
To reduce such background we demand the presence in the range of $|\eta |<3$
of exactly two opposite sign isolated muons with $p_T>7$~GeV. The cut on $p_T$
corresponds to the ATLAS inner tracker performance. The isolation
criterion for the muon is that no cell with $E_T>2$~GeV and no track with
$p_T>7$~GeV is present in a cone of $\Delta R = 0.3$ around the muon track. The
second part of this criterion forces the first one for the central cell of the
cone, whose $E_T$ is ignored.

As shown in Table~1 and in Figs.~16-19, this cut efficiently reduces the number
of events in the first signal sample, but does not effect the
backgrounds from the standard physics processes so much. 
For both signal samples Fig.~21 shows the invariant mass distributions for the
muons 
originating from the same  $\tilde{\chi}_2^0$ in an event. The 
similarity of these distributions demonstrate the similar numbers of decays
(2) in both samples after the application of the muon isolation cut.

The next three cuts reduce background from the standard physics 
processes to a negligible level. They are:

\begin{itemize}
\item the cut on the missing transverse momentum -- $p_T^{miss}>200$~GeV;

\item the cut on the number of jets in the calorimeters -- $n_j > 3$;

\item the cut on the invariant mass of two opposite sign isolated muons --
$m_{\mu \mu} < 80$~GeV.
\end{itemize}

The effect of these cuts is reflected in Table~1 and in Figs.~22-24. From
Figs.~24 and Table~1 one can see that, first, the signal from the gluino
production is clearly seen, and, second, when even all cuts are applied, the
number of events from the decays of $\tilde{g} \ra X$ is almost three times
lager than the number of events from decays (1)-(2). This is caused mainly by
the presence of muons originating from W produced by top quark and chargino
decays.

The composition of the events in the muon transverse momentum distribution 
shown in Fig.~24 for $\tilde{g} \ra X$ decays are as follow:

32.7\% -- $\tilde{g} \ra \tilde{\chi}_2^0 X$, 
$\tilde{\chi}_2^0 \ra \tilde{\chi}_1^0 \mu \mu$;

28.8\% -- $\tilde{g} \ra \tilde{\chi}_2^{\mp} t/\bar t$, 
$\tilde{\chi}_2^{\mp} \ra \tilde{\chi}_1^0 W$, $t \ra Wb$, $W \ra \mu \nu$;

~~7.7\% -- $\tilde{g} \ra \tilde{\chi}_1^{\mp} t/\bar t$, 
$\tilde{\chi}_1^{\mp} \ra \tilde{\chi}_1^0 W$, $t \ra Wb$, $W \ra \mu \nu$;

~~5.8\% -- $\tilde{g} \ra \tilde{\chi}_4^0 X$, 
$\tilde{\chi}_4^0 \ra \tilde{\chi}_1^0 h$, $h \ra \mu \mu$;

15.4\% -- cascade decays in which muons came not from one but from both gluinos;

~~9.6\% -- $\tilde{g} \ra t\bar t \tilde{\chi}^0$, $t \ra Wb$, $W \ra \mu \nu$.

The shapes of the muon $p_T$-distributions in all these channels are very
similar. 
So we can speak about the detection of the muons from different types of
gluino cascade decays. The decays in which the muons originated not from 
cascade gluino decays give a contribution in $p_T$-distribution at the level
of 9.6\% only.

\section*{\large\bf Comparisons with the previous studies}

$E_T^{miss}$ + jets signature was studied in \cite{Polesello} where it was 
shown that to observe a signal from the heavy gluino it is enough to have at
least 3 jets with $E_T > 200$ GeV, a fourth jet with $E_T > 100$ GeV, a
circularity $C > 0.2$, and $E_T^{miss} > 300$ GeV. In \cite{KK} the
isolated muon transverse momentum distributions were investigated for same
sign and for opposite sign dileptons for events with $E_T^{miss} > 100$ GeV. To 
compare our results with both these previous investigations we used our sample
of $\tilde{g} \ra X$ decays (48800 events generated), and a newly generated
sample of 50000 $t\bar t$ production events with all subsequent decays.

Figs.~25 and 26 show $E_T^{miss}$-distributions obtained from central and
forward calorimeter simulations for gluino signal and for $t\bar t$ 
background from those two samples, respectively.

Four histograms are plotted for different sets of cuts in each Figure. 
Here and below P-cuts denote the cuts from
\cite{Polesello} mentioned above. 
Then, for KK-cuts according to \cite{KK} we demand the presence in the
event of $E_T^{miss} > 100$ GeV and at least one either same sign (SS) or
opposite sign (OS) dimuon. In these cases the isolated muons in dimuons have 
transverse momentum greater than 30 GeV and $|\eta |<3$. As
an isolation criterion we used the one described above. Finally, KNS-cuts mean
our cuts described in previous sections of this note.

The numbers of the events shown in the histograms in Figs.~25-26 and also in
Table~2 correspond to the integrated luminosity of $10^5 pb^{-1}$.

\begin{center}
\small

{\bf Table~2} \\
\smallskip
Event rates expected per $10^5 pb^{-1}$ for different cuts \\
\smallskip
\begin{tabular}{|l|r|r|r|r|}    \hline
 & P-cuts & KK-cuts, SS & KK-cuts, OS & KNS-cuts, OS \\ \hline
$\tilde{g} \ra X$ & 14800 &  930 &  2540 &   272  \\ \hline
$t\bar t \ra X$  &  1810 & 2420 & 61000 & $<600$ \\ \hline
\end{tabular}
\end{center}

In a similar way the muon transverse momentum distributions are plotted in 
Figs.~27 and 28. The numbers of muons corresponding to the integrated luminosity
of $10^5 pb^{-1}$ are presented in histograms in these Figures and in Table~3.
An efficiency of 90\% is assumed for muon identification.

\begin{center}
\small

{\bf Table~3} \\
\smallskip
Isolated mion rates expected per $10^5 pb^{-1}$ for different cuts $\dagger$ \\
\smallskip
\begin{tabular}{|l|r|r|r|r|}    \hline
 & P-cuts & KK-cuts, SS & KK-cuts, OS & KNS-cuts, OS \\ \hline
$\tilde{g} \ra X$ &  1900  & 1030 &  3020 &   440  \\ \hline
$t\bar t \ra X$  & $<600$ & 1210 & 85200 & $<600$ \\ \hline
\end{tabular}
\smallskip

$\dagger $ An efficiency of $90\%$ is assumed for muon identification\\
\end{center}

From Tables~2 and 3 one can conclude that the additional demand to detect
exactly two opposite sign isolated leptons in the events gives a good
opportunity to observe a 
signal from the heavy gluino production in $E_T^{miss}$- and in isolated muon
$p_T$-distributions. The numbers we obtained applying P-cuts or KK-cuts to
our samples are in consistence with the previous studies \cite{Polesello, KK}.

\section*{\large\bf Conclusions and further steps}

We conclude, therefore, that the combination of $E_T^{miss}$ + jets cuts with
the detection of exactly one 
opposite sign dilepton is very promising signature to reduce the 
backgrounds to the heavy gluino pair production. 
The cuts applied to the decays of $\tilde{g} \ra X$ and to the background events
are enough to extract a SUSY signal over Standard Model background.  They are
however not enought to extract a 
signal from the cascade decays (1)-(2) from other heavy gluino decays. 
A possible improvement could be found by better cut optimizations and more
careful investigation of the muon isolation in both signal samples. Another way
could be to tag b-quark from the 
decay of $t \ra W b$. The results obtained in present paper can also change
when the pile-up effect and muon momentum resolution will be included 
in consideration. 

\bigskip

{\em Acknowledgements.} The authors are thankful to P.~Jenni, D.~Froidevaux and
S.~Hellman for the support of this work and prolific discussions.

\includegraphics{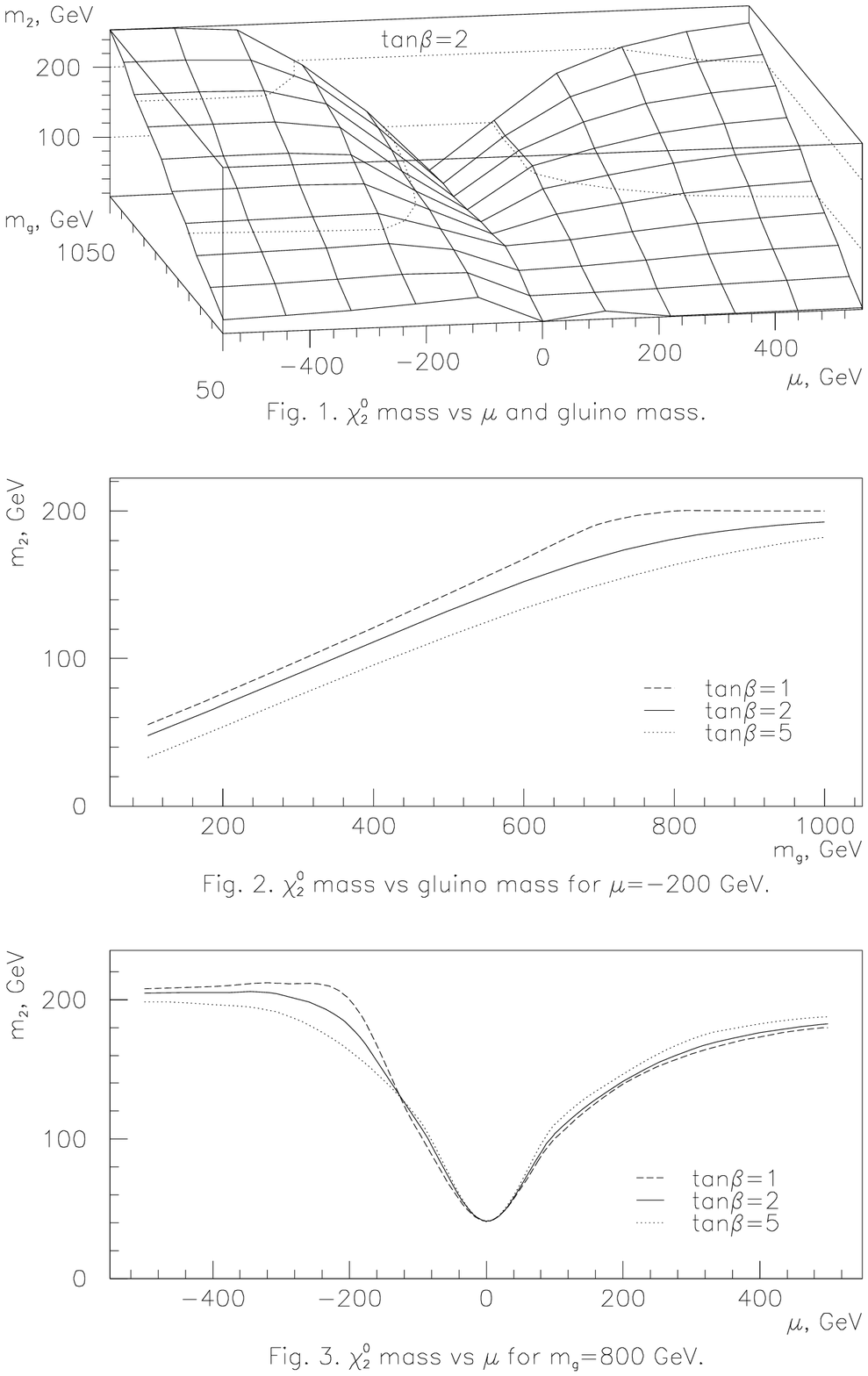}
\newpage
\includegraphics{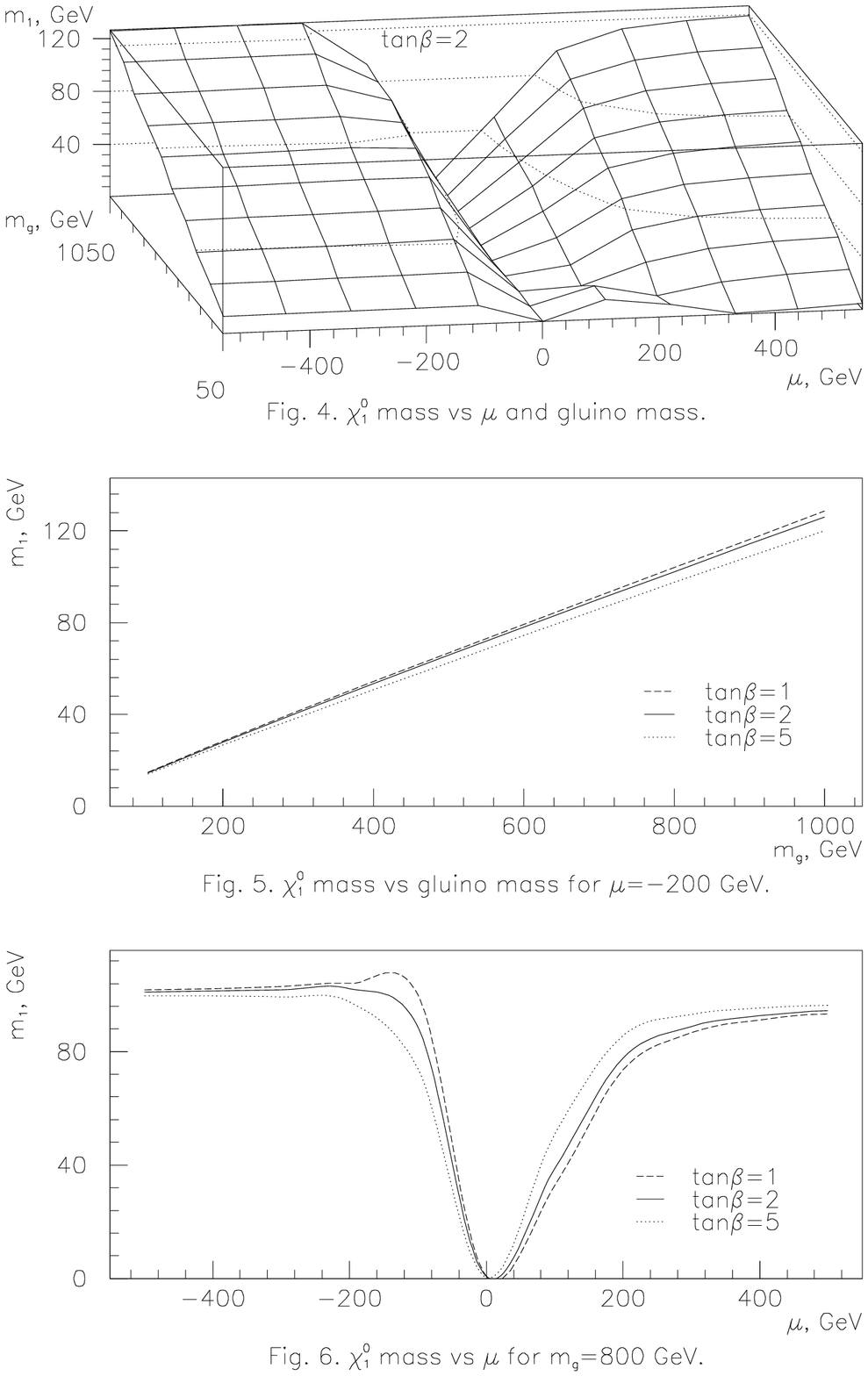}
\newpage
\includegraphics{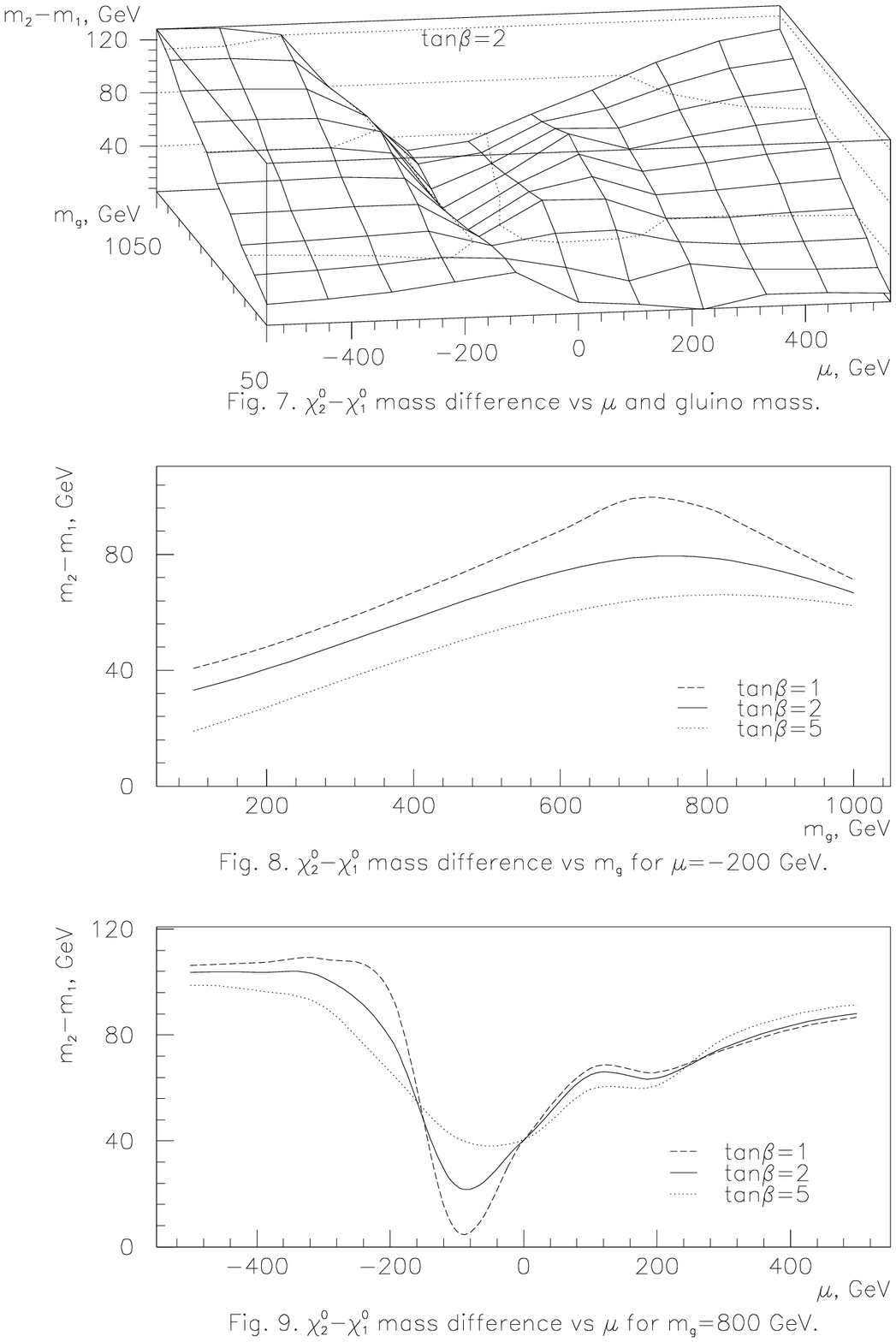}
\newpage
\includegraphics{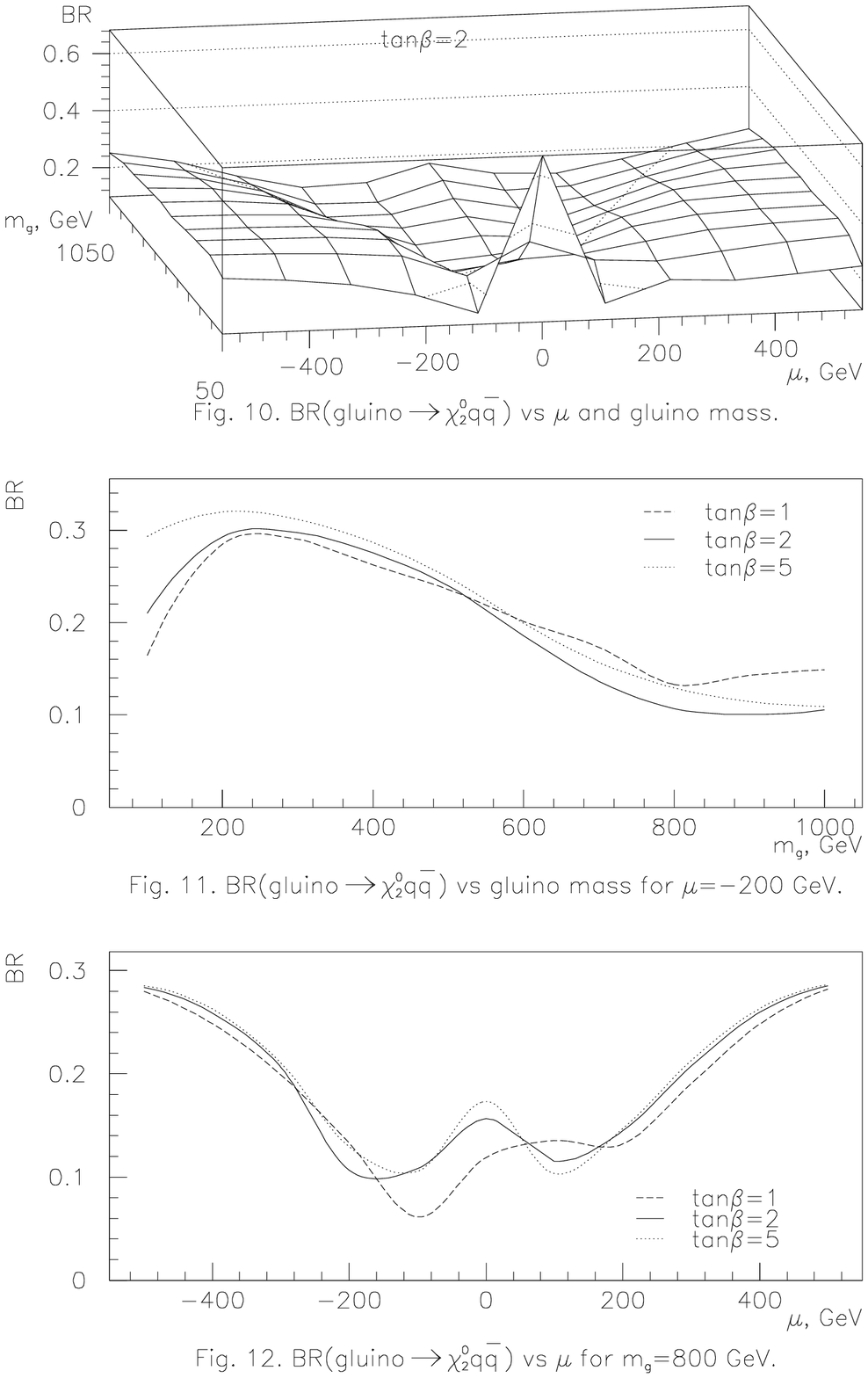}
\newpage
\includegraphics{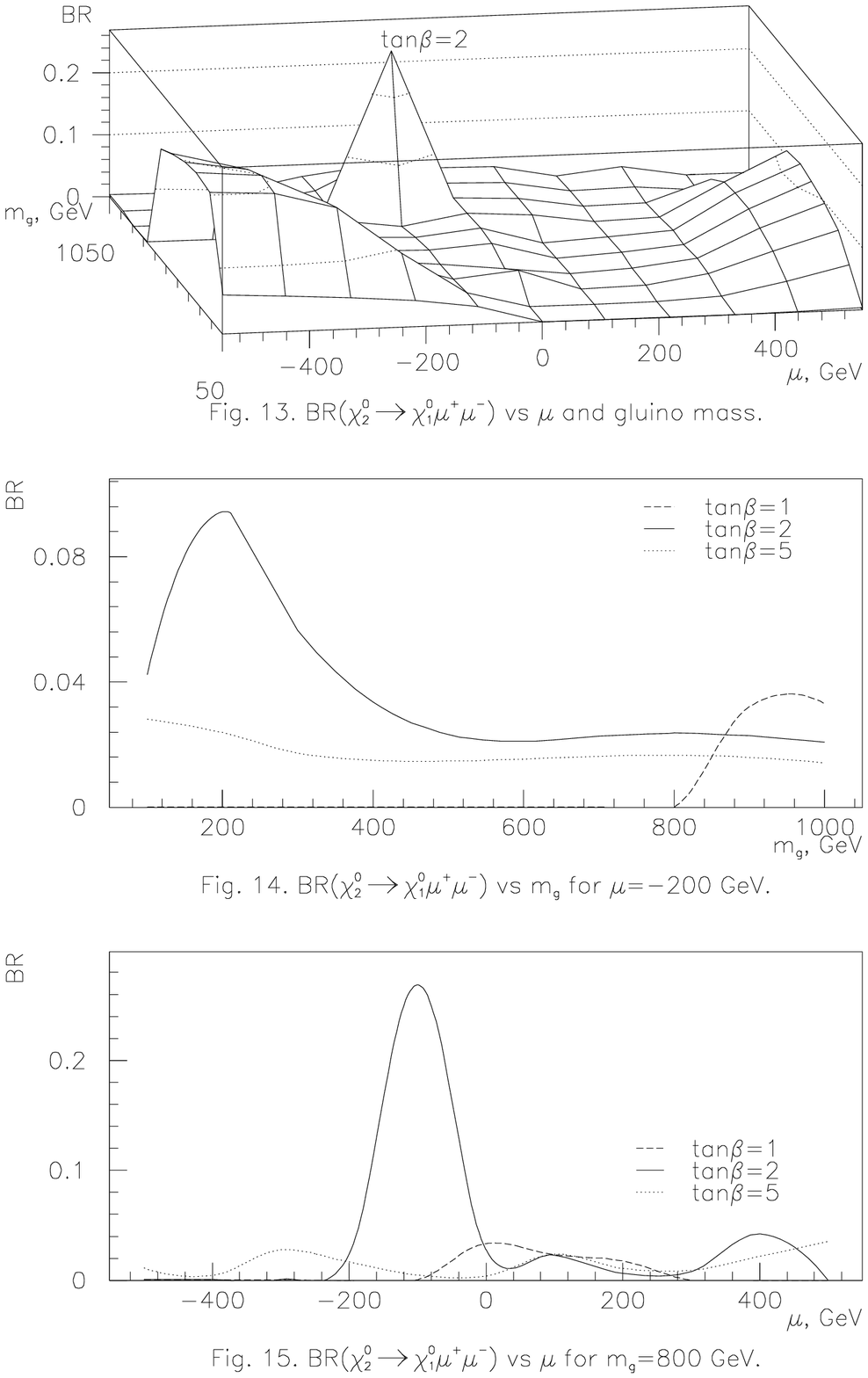}
\newpage
\includegraphics{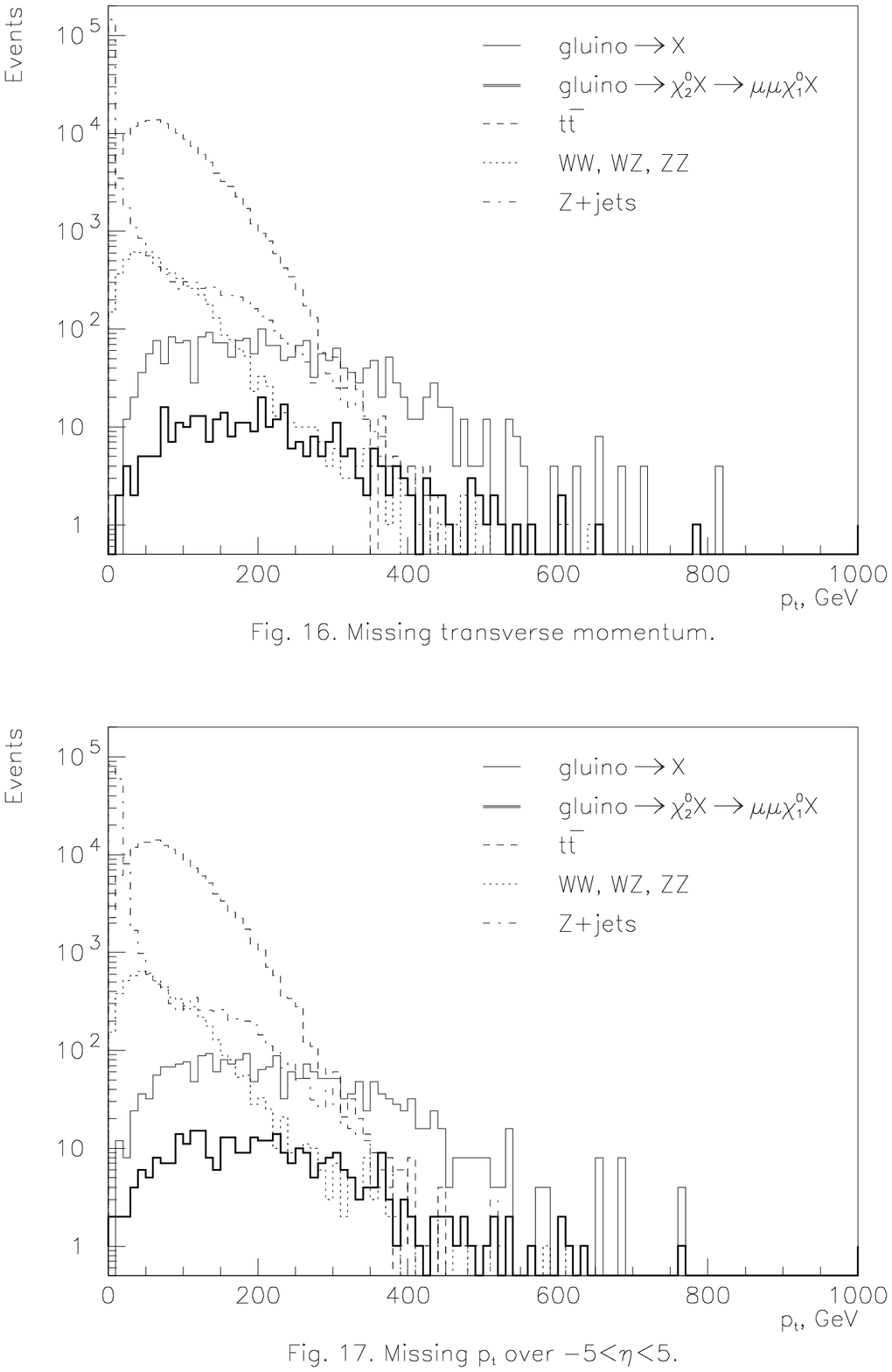}
\newpage
\includegraphics{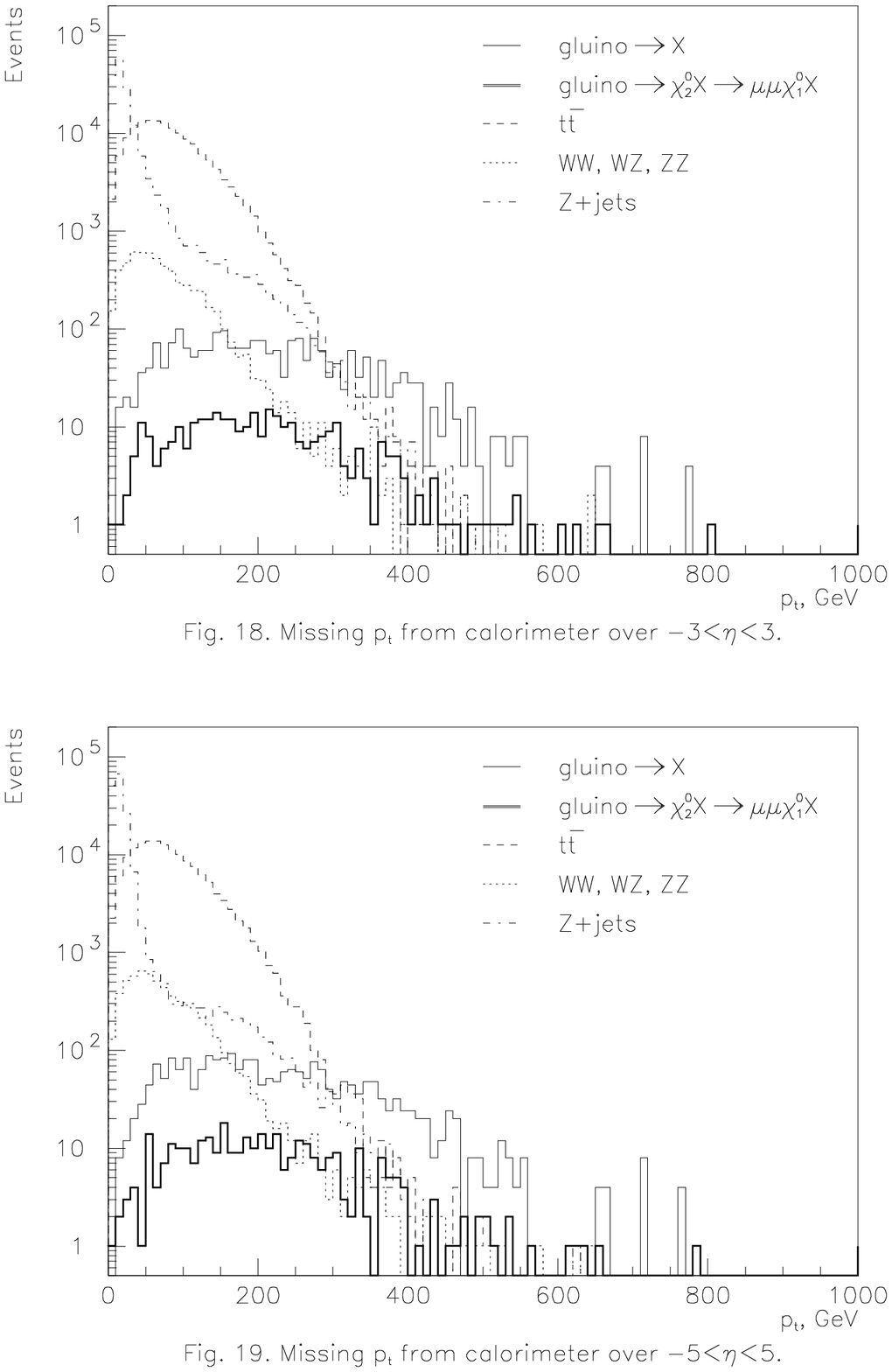}
\newpage
\includegraphics{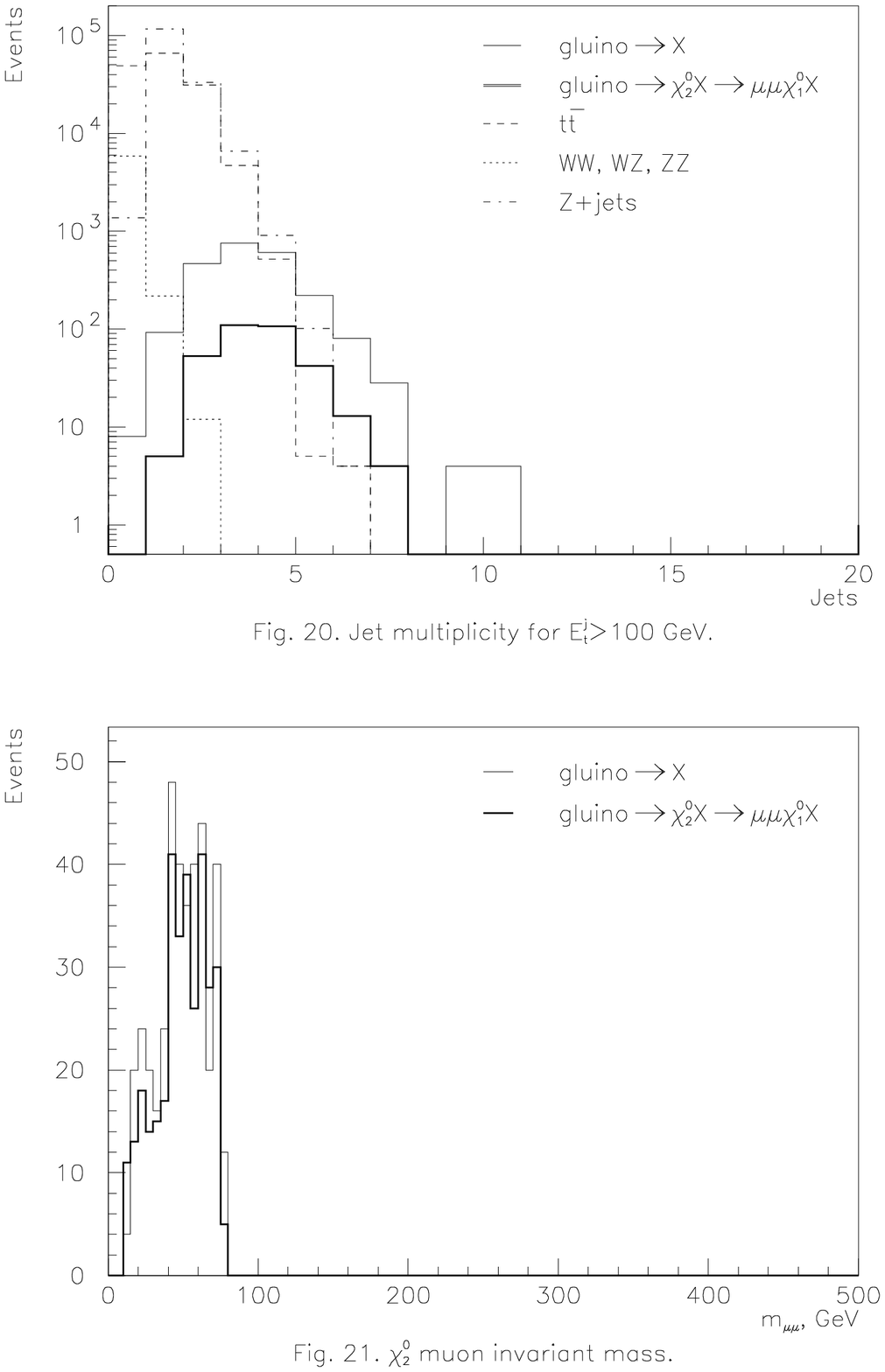}
\newpage
\includegraphics{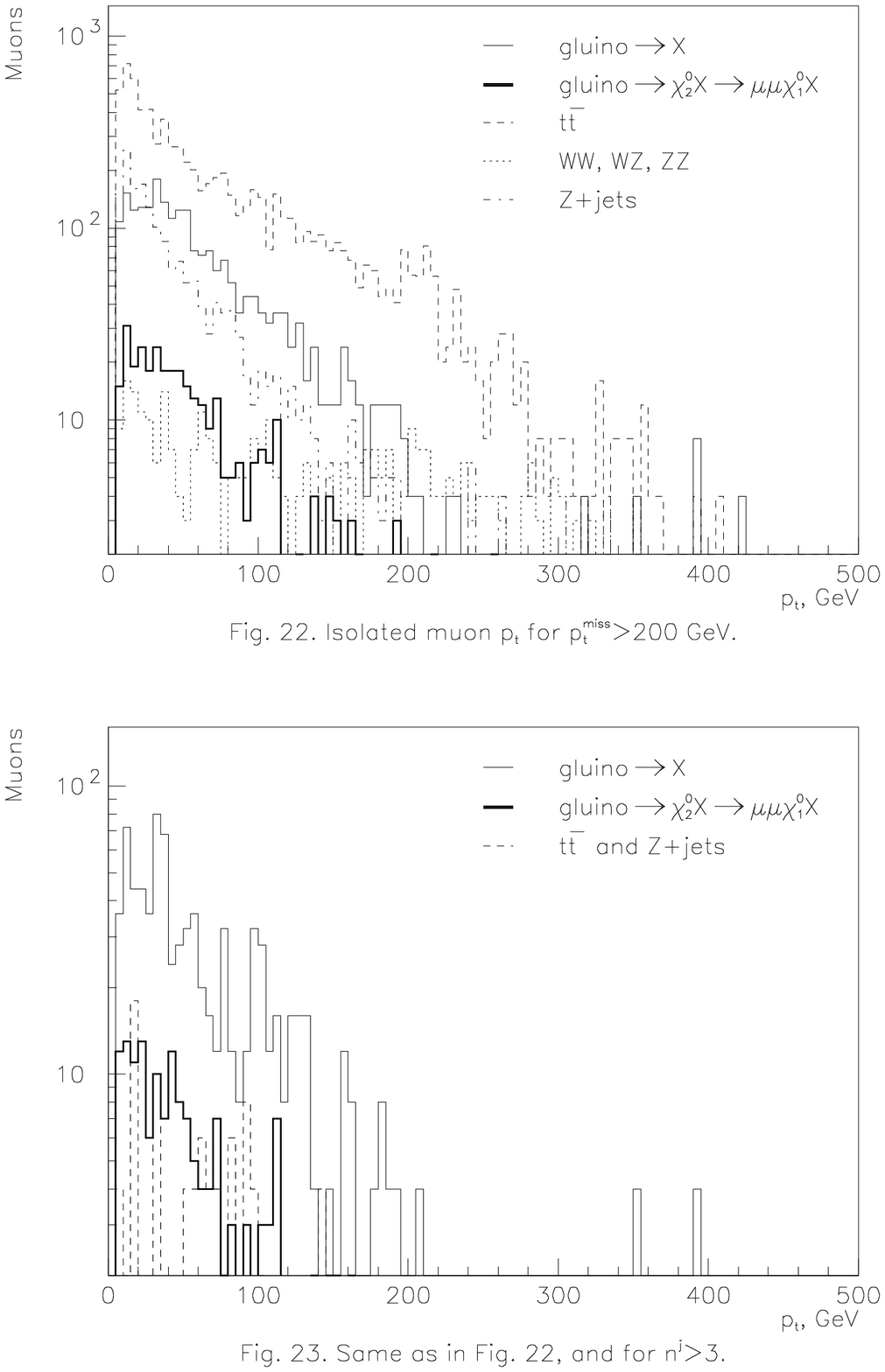}
\newpage
\includegraphics{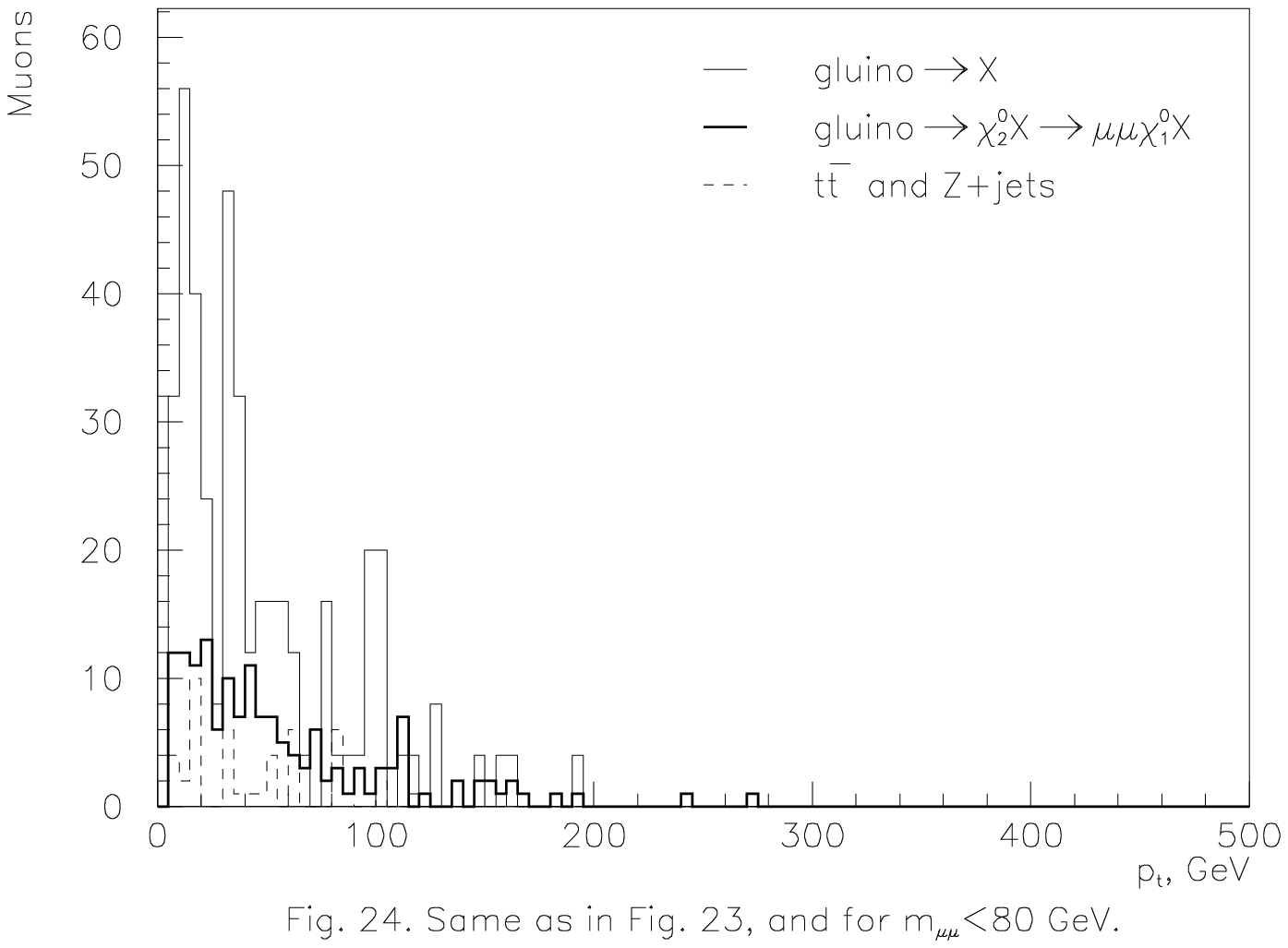}
\newpage
\includegraphics[width=\textwidth]{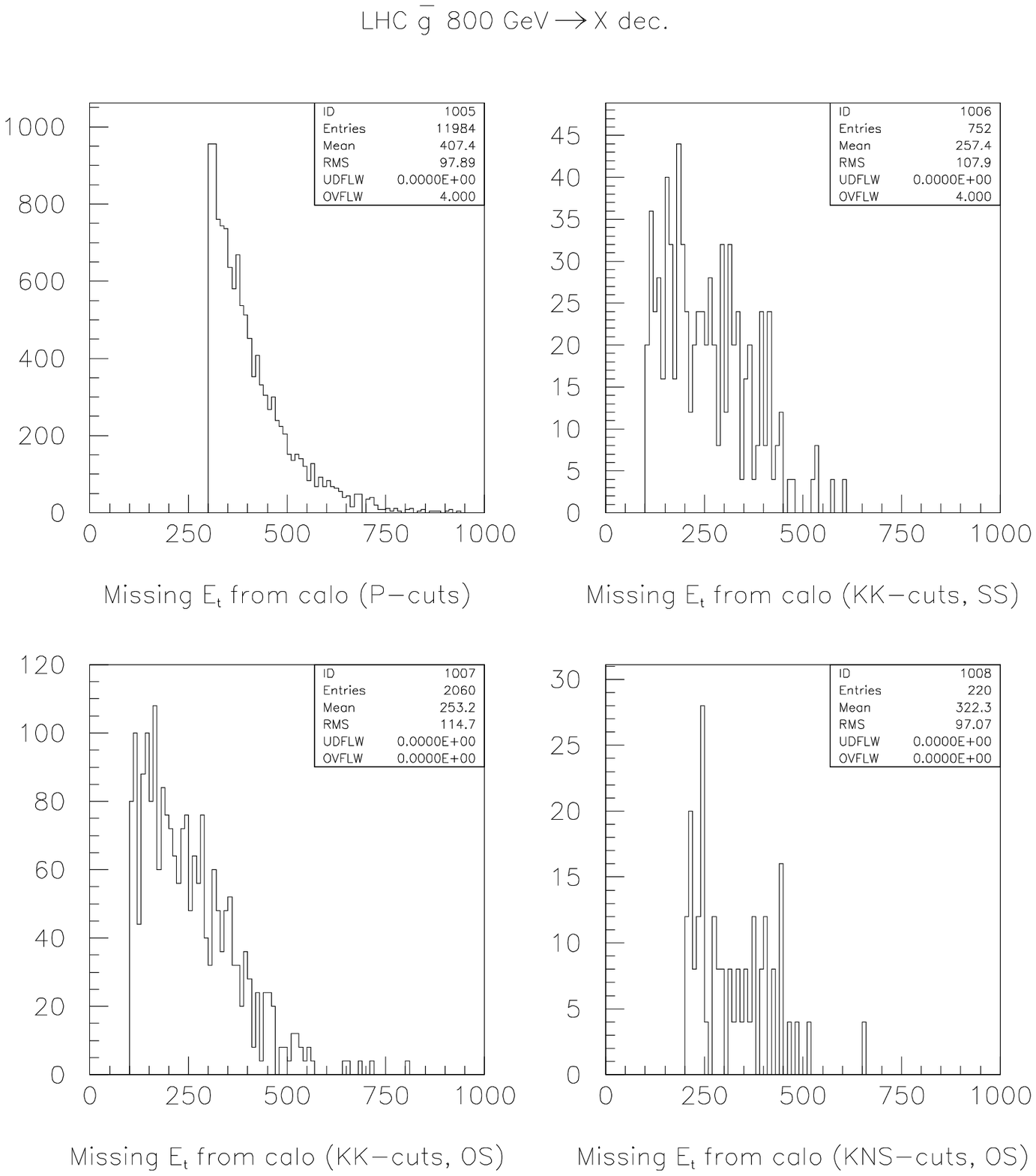}

\begin{center}
Fig. 25. Event missing transverse energy distribution for gluino signal for different sets of cuts.
\end{center}
\newpage
\includegraphics[width=\textwidth]{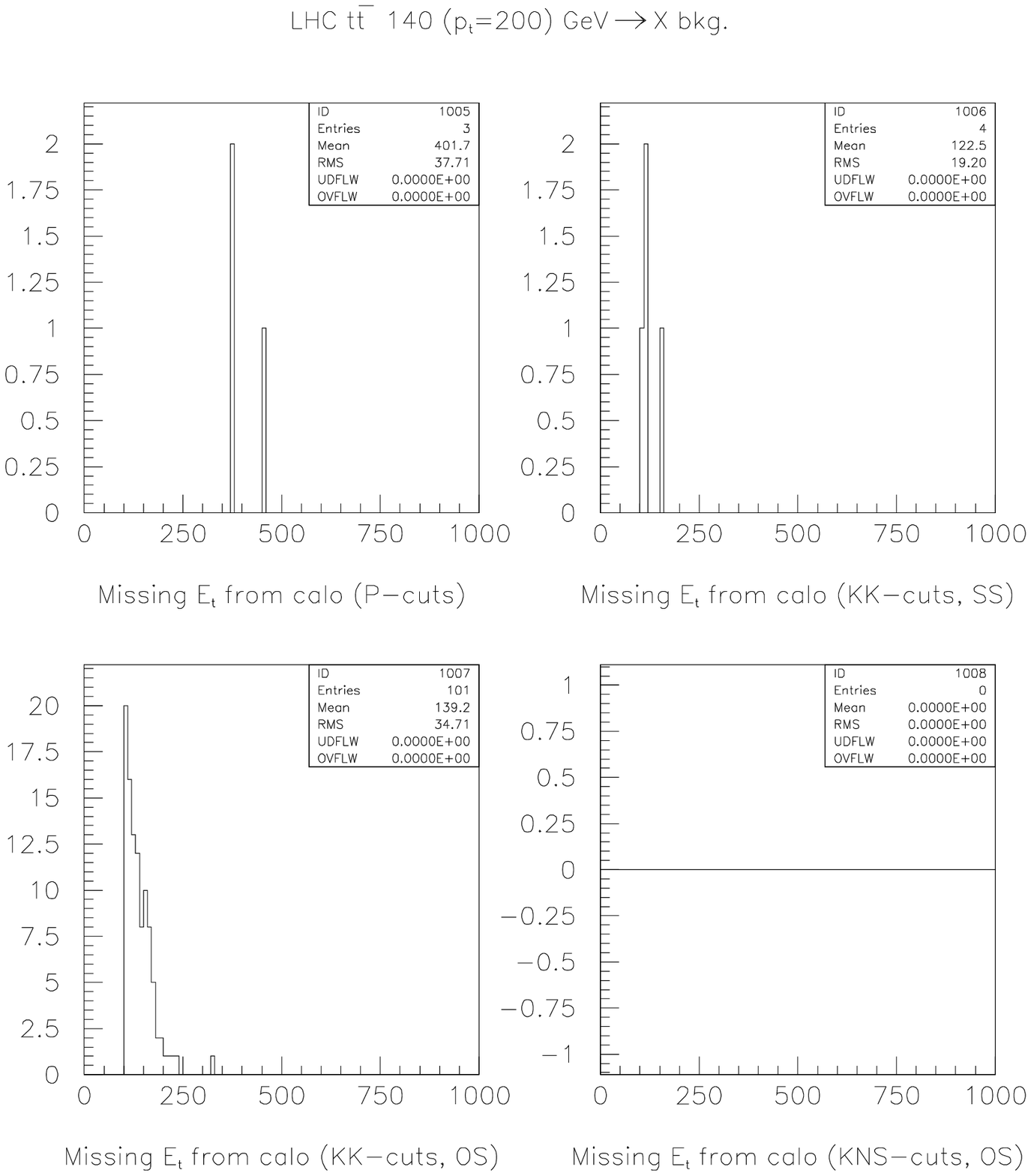}

\begin{center}
Fig. 26. Event missing transverse energy distribution for $t\bar t$ background for different sets of cuts.
\end{center}
\newpage
\includegraphics[width=\textwidth]{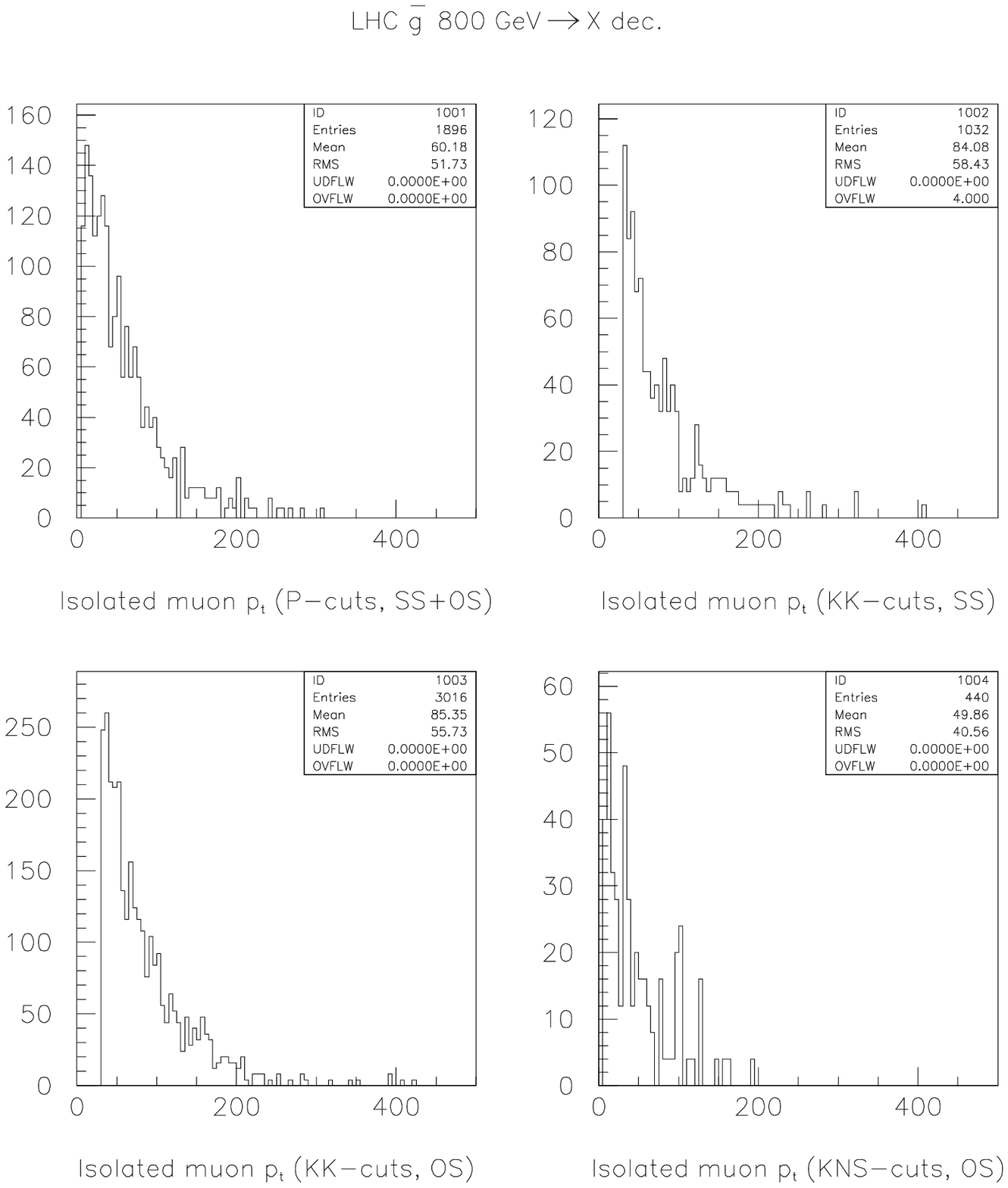}

\begin{center}
Fig. 27. Event missing transverse momentum distribution for gluino signal for different sets of cuts.
\end{center}
\newpage
\includegraphics[width=\textwidth]{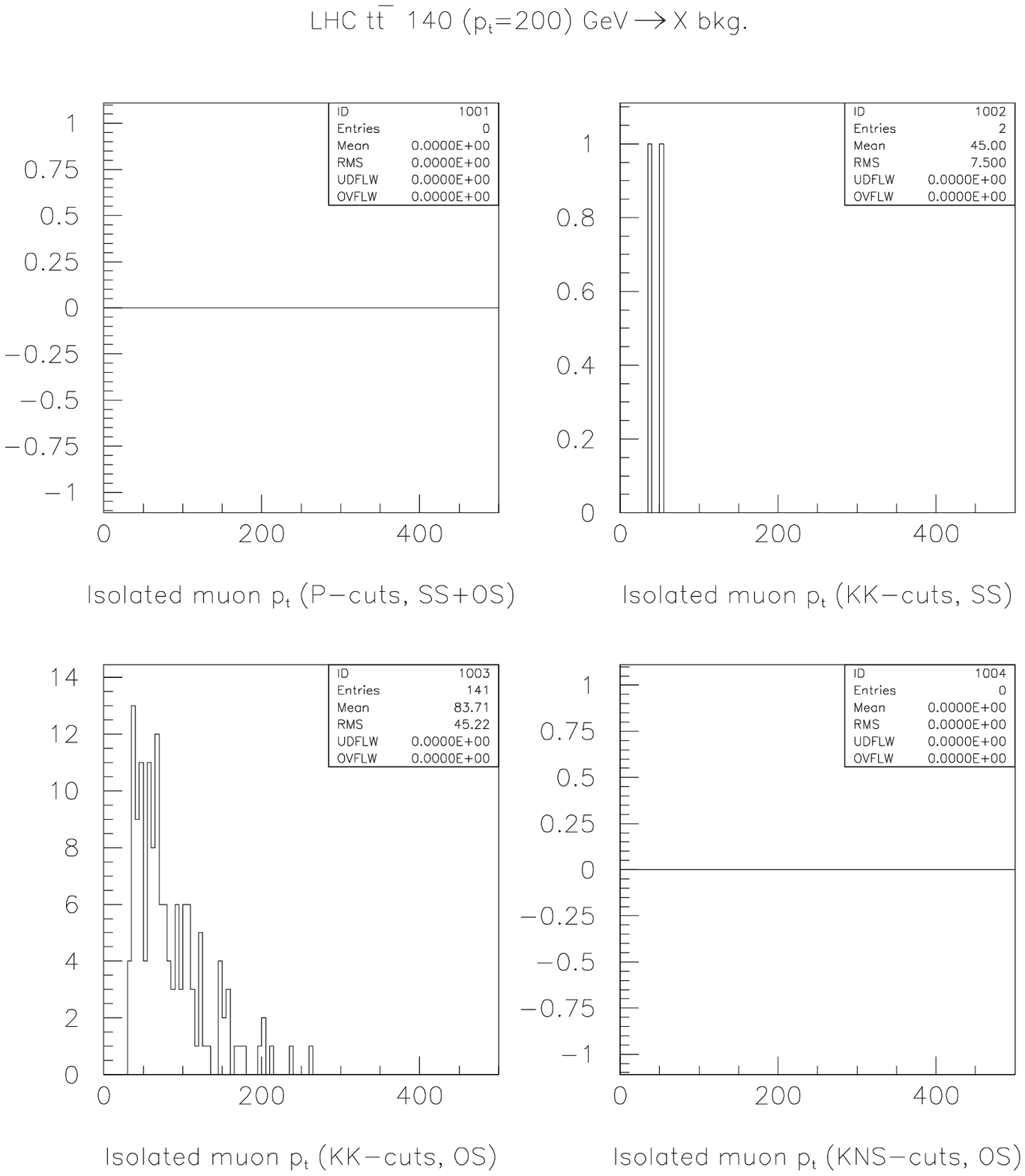}

\begin{center}
Fig. 28. Event missing transverse momentum distribution for $t\bar t$ background for different sets of cuts.
\end{center}


\begin{thebibliography}{99}
\bibitem{ATLAS}
ATLAS Letter of Intent, CERN/LHCC/92-4, LHCC/I 2, Geneva, 1992.
\bibitem{Sokolov}
A.\,A. Neushkin, A.\,A. Sokolov, G.\,G. Volkov,
{\em Search for SUSY particles at the UNK energies.} 
IHEP Preprint 92-70, Protvino, 1992.
\bibitem{ISAJET}
ISAJET 6.50 written by F.\,E. Paige and S.\,D. Protopopescu.
\bibitem{ISASUSY}
ISASUSY 1.1 written by H.\,Baer, F.\,E. Paige, S.\,D. Protopopescu, 
and X. Tata.
\bibitem{MSSM}
H.\,P. Nilles, Phys. Rep. 110 (1984) 1; \\
G. Ridolfi, G.\,C. Ross, F. Zwirner, in Proceedings of Large Hadron Collider
Workshop, Aachen 4-9 October 1990, edited by G. Jarlskog and D. Rein, CERN 
90-10, ECFA 90-133, Vol. II, p. 608. 
\bibitem{Polesello}
G. Polesello, {\em Search for gluino pair production in the $E_T^{miss}$
+ jets channel.} ATLAS Internal Note PHYS-No-16, 29 January 1993.
\bibitem{KK}
K. Kawagoe and S. Komamiya, {\em Search for gluino pair production trough 
their cascade decay into same sign dileptons.} ATLAS Internal Note 
PHYS-No-14, 13 January 1993.

\end{thebibliography}
\end{document}